# Non-monotonic Fermi surface evolution and its correlation with stripe ordering in bilayer manganites

Z. Sun[1], Q. Wang[1], J. F. Douglas[1], Y.-D. Chuang[2], A. V. Fedorov[2], E. Rotenberg[2], H. Lin[3], S. Sahrakorpi[3], B. Barbiellini[3], R. S. Markiewicz[3], A. Bansil[3], H. Zheng[4], J. F. Mitchell[4], and D. S. Dessau[1]

[1] *Department of Physics, University of Colorado, Boulder, CO 80309, USA*

[2] *Advanced Light Source, Lawrence Berkeley National Laboratory, Berkeley, CA 94720, USA*

[3] *Department of Physics, Northeastern University, Boston, MA 02115, USA*

[4] *Materials Science Division, Argonne National Laboratory, Argonne, IL 60439, USA*

In correlated electron systems such as cuprate superconductors and colossal magnetoresistive (CMR) oxides there is often a tendency for a nanoscale self-organization of electrons that can give rise to exotic properties and to extreme non-linear responses. The driving mechanisms for this self-organization are highly debated, especially in the CMR oxides in which two types of self-organized stripes of charge and orbital order coexist with each other. By utilizing angle-resolved photoemission spectroscopy measurements over a wide doping range, we show that one type of stripe is exclusively linked to long flat portions of nested Fermi surface, while the other type prefers to be commensurate with the real space lattice but also may be driven away from this by the Fermi surface. Complementarily, the Fermi surface also appears to be driven away from its non-interacting value at certain doping levels, giving rise to a host of unusual electronic properties.

Strong electron correlation effects of various kinds can yield nanoscale self organizations of charges, spins, and orbitals, often forming stripe-like patterns (*1,2,3*). These patterns are generally believed to be highly relevant for the exotic physical properties — in the manganites for example the temperature-dependent behavior of their scattering intensity typically resembles that of the resistivity, arguing for a causal relationship (see Fig. 1D). Despite their importance, little direct information exists about how stripe modulations alter or are altered by the electronic band structure. More generally, the manner in which correlation effects alter the non-interacting (e.g. band theory) electron behavior is at the heart of modern condensed matter physics, though is poorly understood both from theoretical and experimental grounds. In most cases, the electronic states near the Fermi level have modified slopes (velocities) and curvature (masses), but otherwise the Fermi surface parameters remain essentially unchanged and can be predicted from band theory. However, in the presence of the self-organization of electrons the reliability of the well-established band theory encounters a greater challenge, with the study of this topic highly relevant to many problems in modern condensed matter physics.

In the family of bilayer manganites $La_{2-2x}Sr_{1+2x}Mn_2O_7$, the delicate competition and interplay among charge, spin and orbital degrees of freedom varies with doping and yields a wide diversity of phenomena as shown in the doping phase diagram of Fig. 1A (*4,5*). Some of the important regimes such as the CE insulator at $x = 0.50$ and the bi-stripe insulator at $x = 0.60$ are explicitly related to a diagonal stripe type of charge/orbital order (figure 1B), though these stripes extend to other parts of the phase diagram with short-range order (*5,6,7,8,9,10,11*). There also is a vertical type of stripe order (Fig 1C) which exists with short range order for $x < 0.5$ (*12,13,14,15*). These stripes have gotten lots of attention, partly because the scattering intensity has a temperature dependence that very closely matches the electrical resistivity (Fig 1D) (*6,7,8,9,12,13,16*), indicating their relevance for the transport properties of these materials. The origin and nature of these stripe modulations are, however, still quite controversial.

In *k*-space (Fig. 1E), the Fermi surface is dominated by one electron-like pocket (black) centered at the *Γ* point and two hole-like pockets (green) around the zone corners (*17,18,19*). The straight segments near the zone boundary are the bonding (solid green) and the antibonding (dashed green) portions of Fermi surface of bilayer-split bands that arise from the coherent hopping between the two neighboring $MnO_2$ planes per unit cell (*20*). The bonding portions of the Fermi surface are straighter than the antibonding ones, so we expect these states to be more important for any ordering tendencies — an expectation which is borne out in our experiments. Superimposed on the Fermi surface are *q*-vectors for both the diagonal or D stripes (blue) and the vertical or V stripes (red). It is seen that both of them have a possibility to nest the straight bonding Fermi surface portions, though whether this actually will happen is under debate (*21,22*). ARPES is uniquely suited to measure these Fermi surface *k*-vectors. As will be shown later, the correlation between stripes and Fermi surface nesting depends upon the specific stripe *q*-vectors and Fermi surface *k*-vectors, which in turn depends upon the doping levels.

Fig. 2A shows ARPES momentum distribution curves (MDCs) at the Fermi level taken along the black cut in Fig. 1E (antinodal states) over a large doping range. Different photon energies were used to emphasize the most relevant portion of the data (*20*), that is, the bonding bands (in the presence of bi-layer splitting band (*20*)) or the degenerate band (when the splitting between bonding and antibonding bands disappears, which occurs at higher doping levels (*19*)). The separation of the double peaks is a measure of the nesting vectors. These nesting vectors, $2k_F$, are plotted in Fig. 2B (green circles) as a function of doping level. They have a large value $2k_F \sim 0.30 \times (2\pi/a, 0)$ for intermediate dopings $x \sim 0.40 - 0.50$, with noticeably smaller values at both higher and lower dopings. Such a non-monotonic evolution with doping is unexpected from electronic structure theory (red hatched region - up triangles (see supporting material) and down triangles (*22*)). In general the theoretical and experimental data show an excellent match for $x > 0.5$, both including the overall magnitudes and the slope of the trends. For $0.4 < x < 0.5$ the overall magnitudes of the vectors are roughly comparable but we notice that the doping trend is different — the experimental data shows a nearly flat trend with reduced doping while the theory clearly continues to increase. This disagreement accelerates even more dramatically for $x < 0.4$ where the trend in the experimental data reverses. Such an unusual behavior, breaking from electronic structure theory, speaks of novel physics. First, however, we must argue that the observed effect is representative of the bulk. This can be convincingly demonstrated by a comparison of the *k*-vectors measured in our ARPES experiments to *q*-vectors measured in bulk-sensitive x-ray and neutron scattering experiments, as will be discussed below. The clear correlation we will show between these different types of experiments over so large a doping range indicates that the Fermi surfaces measured by the surface-sensitive ARPES technique are representative of the bulk, since they agree so well with the bulk-sensitive scattering data.

Using ARPES, Chuang *et al.* first pointed out the correlation between the vertical stripes and the Fermi surface nesting (bonding band), for the $x = 0.4$ sample they studied had a separation in *k*-space very similar to the scattering vectors of vertical stripes (*21*). This agreement can be seen in Fig. 2B, where for $x = 0.4$ both the green circle (ARPES $2k_F$) and the red V (*q* vector from x-ray scattering) have the same magnitude $\sim 0.3 \times (2\pi/a)$ (*15,21*). More recently however, it has been called into question whether the vertical stripe vector is connected with the Fermi surface nesting behavior because it was observed that the magnitude of the vertical stripe vectors from the scattering experiments varies with doping in the opposite direction (*12*) to the anticipated evolution of the Fermi surface by band structure calculations (*22*). This is most obvious for $x < 0.4$ where the V vectors decrease with smaller $x$, while the band calculations have an increasing value (red hatched region). One of the new results here, unexpected from previous experience, is that the Fermi surface nesting vectors $2k_F$ (green circles) from ARPES measurements also have a backwards trend from the band calculations. In fact, Fig. 2B shows that the observed V vector very closely matches the measured Fermi surface nesting vector $2k_F$ over the entire range where this comparison can be made. For $x < 0.5$, an x-ray scattering experiment with even finer doping steps has been performed (*23*) which yields a result highly consistent with our ARPES measurements. This is, we argue, conclusive evidence that the V-stripe charge/orbital modulations are intrinsically locked to, and originate from, Fermi surface nesting. In this respect the V-stripe modulations appear very similar to a classic charge density wave (CDW) and are driven by the nesting properties of the Fermi surface, although as we will discuss later here they are richer — consisting of an unusual temperature dependence, very high energy scales, *etc*.

We have also noticed a similar connection between the D vectors and the ARPES $2k_F$ for certain doping levels, though not for all doping levels. The D-type modulation possesses one well-understood fixed point at the commensurate doing level $x = 0.5$, where the material is a charge-ordered antiferromagnetic insulator. Here, it is termed "CE" and has a beautiful real-space picture (Fig. 1B) originally due to Goodenough in 1955 (*24*), which includes the modulation of the charge, spin, orbital, and lattice degrees of freedom. This real-space modulation corresponds to a relatively short vector in *k*-space: $q = (0.25, 0.25, 0)$ in units of $2\pi/a$. Away from $x = 0.5$ the long-range commensurate modulation of the D vector is lost — in the underdoped regime ($x = 0.4 – 0.5$) it keeps its magnitude but becomes short range, while in the overdoped regime the D vector magnitude decreases noticeably (blue D's in Fig. 2B). In the regime of $x = 0.4 – 0.5$, the D vectors are far away from the nesting vectors but remain commensurate with the underlying main crystal lattice, indeed suggesting a real-space origin of these modulations. When the D vectors deviate from ((0.25,0.25,0)) outside the range of $x = 0.4 – 0.5$ doping, we find that they closely match the Fermi surface nesting vectors $2k_F$. This is apparent between $x = 0.5$ and $0.6$ as well as for the one point $x = 0.38$ in the underdoped region. Apparently the D vectors have two choices — either to lock in commensurately with the real space lattice or to slightly alter the magnitude so as to nest the Fermi surface, which indicates that D stripes are under a dual influence from both the lattice structure and the Fermi surface topology.

An especially interesting doping point on the diagram is $x = 0.38$, where we see that the scattering vector is pulled away from the commensurate D value of $q = 0.25$ to $q = 0.27 \times 2\pi/a$. Here we also see that the Fermi surface nesting vector has been pulled down by a significant amount from its "expected", i.e. band calculation value, to the nesting vector $(2k_F, 2k_F) \times (\pi/a)$ with $k_F = 0.27$ which exactly matches the D vector periodicity. The agreement between these two values is unlikely to be a coincidence and suggests strong positive feedback — the ability to nest the Fermi surface pulls the D vector away from its preferred value, while also pulling the Fermi surface from its preferred value. To our knowledge, this is the first clear evidence in the literature of a Fermi surface crossing being modified so as to more positively nest a charge/orbital modulation. Measurements on the related copper-oxychlorides also showed a connection between real-space charge modulations and structures in *k*-space (*25*). In that case however, the relevant ARPES features were so broad and incoherent, and the dispersion so vertical, that it was not even clear if the term "Fermi surface" had meaning. This is not the case for the $x = 0.38$ manganites which have strong and sharp quasiparticle excitations at the Fermi level. Such a feedback effect on the Fermi surface from the ordering vectors is a beautiful example of the coupling of the spin, charge, and orbital degrees of freedom. Regarding the feedback between Fermi surface nesting and charge/orbital ordering, one may wonder how the (static) D stripes, which have so far been observed only above $T_c$, can send a feedback to the Fermi surface topology measured at low temperature. We note that an indirect evidence of the dynamic CE stripes below $T_c$ in $x = 0.4$ samples has been reported by Weber *et al.* (*26*). This gives rise to an interesting possibility — the existence of dynamic stripes below $T_c$ would allow the Fermi surface nesting and charge/orbital ordering to interact at low temperatures, and they could reach a characteristic vector compatible with each of them.

We also note there are clear difference in other physical parameters of the samples with $x < 0.39$ compared to those with $x = 0.40$ or greater (*20,27*). In particular, by observing a metallic Fermi edge above $T_c$ where the samples are globally insulating, we recently showed the emergence of electronic phase separation into metallic and non-metallic regions for $x < 0.39$ (*27*), which is likely equivalent to phase separation into hole-rich and hole-poor regions. This evidence for phase separation is not apparent in samples with $x > 0.4$ (*21, 28*). Such a phase separation may be one of nature's ways of allowing the electron count to match the chemical doping level, even while the Fermi surface is distorted away from Luttinger's homogeneous value.

Scattering measurements indicate that the V stripes exist up to the order of 500K (*23*). From this we might imagine a BCS weak coupling CDW gap $\Delta = 1.76 k_B T_c \sim 75$ meV, with the main dispersive band reaching this energy scale. In contrast, we find gaps on the scale of hundreds of meV as well as broad EDC peaks which are centered at even higher binding energies (Fig. 3A for an example from the $x = 0.4$ material), indicating the cooperation of the stripe/CDW physics with another higher energy scale such as polarons, Mott, or orbital physics (*17,28*). The large energy scale gaps the entire Fermi surface (Fig. 3C) and drives the material insulating, though there is still a clear underlying *k*-dependence at deeper energies (Fig. 3D) implying that the electrons are still delocalized Bloch-like states. At low temperature, where the energy gain from electron itinerancy driven by double-exchange physics (*29,30*) begins to take over, the stripes dissolve and electronic states begin to leak into the gaps (Figs 3B, E, and F).

The unusual doping evolution of the Fermi surface in $La_{2-2x}Sr_{1+2x}Mn_2O_7$ sheds light on electron correlation effects, and the failure of band theory calls for a more comprehensive understanding of novel materials. The present work brings the origins of different stripes to our concern, strengthens the primary role played by the Fermi surface topology, and also indicates a very large energy scale for the relevant physics. These findings are directly relevant to some of the major questions in many of today's most interesting systems, in which a nanoscale self-organization of the charge, spin, and orbital degrees of freedom has also been observed (*25,31*).

---

1 J. Zaanen, *Science* **286**, 251 (1999).

2 V. J. Emery, S. A. Kivelson, J. M. Tranquada, *Proc. Natl. Acad. Sci.* **96**, 8814 (1999).

3 E. Dagotto, Science 309, 257 (2005).

4. J. F. Mitchell *et al.*, *J. Phys. Chem. B* **105**(44), 10731 (2001).

5. H. Zheng, Q. A. Li, K. E. Gray, J. F. Mitchell, *Phys. Rev. B* **78**, 155103 (2008).

6 Q. A. Li *et al.*, *Phys. Rev. Lett.* **98**, 167201 (2007).

7. T. Kimura, R. Kumai, Y. Tokura, J. Q. Li, Y. Matsui, *Phys. Rev. B* **58**, 11081 (1998).

8. Q. A. Li *et al.*, *Phys. Rev. Lett.* **96**, 087201 (2006).

9. J. Q. Li, C. Dong, L. H. Liu, Y. M. Ni, *Phys. Rev. B* **64**, 174413 (2001).

10. D. N. Argyriou *et al.*, *Phys. Rev. Lett.* **89**, 036401 (2002).

11. T. A. W. Beale *et al.*, *Phys. Rev. B* **72**, 064432 (2005).

12. M. Kubota *et al.*, *J. Phys. Soc. Jpn.* **69**, 1986 (2000).

13. L. Vasiliu-Doloc *et al.*, *Phys. Rev. Lett.* **83**, 4393 (1999).

14. L. Vasiliu-Doloc *et al.*, *J. App. Phys.* **89**, 6840 (2001).

15 B. J. Campbell *et al.*, *Phys. Rev. B* **65**, 014427 (2001).

16 J. F. Mitchell *et al.*, *Phys. Rev. B* **55**, 63 (1997).

17. D. S. Dessau *et al.*, *Phys. Rev. Lett.* **81**, 192 (1998).

18. R. Saniz, M. R. Norman, A. J. Freeman, *Phys. Rev. Lett.* **101**, 236402 (2008).

19. Z. Sun *et al.*, *Phys. Rev. B.* **78**, 075101 (2008).

20. Z. Sun *et al.*, *Phys. Rev. Lett.* **97**, 056401 (2006).

21. Y. D. Chuang, A. D. Gromko, D. S. Dessau, T. Kimura, Y. Tokura, *Science* **292**, 1509 (2001).

22. M. W. Kim *et al.*, *Phys. Rev. Lett.* **98**, 187201 (2007).

23. S. Rosenkranz, private communication

24 . J.B. Goodenough,. Theory of the role of covalence in the perovskite-type manganites [La,M(II)]MnO$_3$. *Phys. Rev.* **100**, 564 (1955).

25 K. M. Shen *et al.*, *Science* **307**, 901 (2005).

26. F. Weber *et al.*, *Nature Physics* **8**, 798 (2009).

27. Z. Sun *et al.*, *Nature Physics* **3**, 248 (2007).

28. N. Mannella *et al.*, *Nature* **438**, 474 (2005).

29. P.W. Anderson, H. Hasegawa, *Phys. Rev.* **100**, 675 (1955).

30. P.-G. de Gennes, *Phys. Rev.* **118**, 141 (1960).

31. W.D. Wise *et al.*, *Nature Phys.* 4, 696 (2008).

32. We thank T. Devereaux, D. Reznik, D. N. Argyriou and S. Rosenkranz for helpful discussions. This work was supported by the U.S. National Science Foundation under grant DMR 0706657. The Advanced Light Source is supported by the Director, Office of Science, Office of Basic Energy Sciences, of the U.S. Department of Energy under Contract No. DE-AC02-05CH11231. The theoretical work is supported by the US Department of Energy contracts DE-FG02-07ER46352 and DE-AC03-76SF00098, and benefited from the allocation of supercomputer time at the NERSC and the Northeastern University's Advanced Scientific Computation Center (ASCC). Argonne National Laboratory, a U.S. Department of Energy Office of Science Laboratory, is operated under Contract No. DE-AC02-06CH11357. The U.S. Government retains for itself, and others acting on its behalf, a paid-up nonexclusive, irrevocable worldwide license in said article to reproduce, prepare derivative works, distribute copies to the public, and perform publicly and display publicly, by or on behalf of the Government.

Correspondence and requests for materials should be addressed to D.S.D. (e-mail: Dessau@colorado.edu) or Z.S (SunZhe@gmail.com)

Fig. 1

Characteristics of $La_{2-2x}Sr_{1+2x}Mn_2O_7$. (**A**) A schematic phase diagram of $La_{2-2x}Sr_{1+2x}Mn_2O_7$. The complicated interactions in this material yield a diversity of metallic, insulating, paramagnetic, ferromagnetic (FM), antiferromagnetic (AF) states. At exactly x=0.50 and x=0.60, the material is an insulator with CE-type and bi-stripe long range charge/orbital ordering, respectively. White dots indicate temperature and doping levels considered in this paper. (**B**) "CE" ordering in real space, which is a specific case of diagonal stripes. (**C**) Short range (0.3, 0, 1) vertical stripes in x=0.4 compound, after

ref.*15*. (**D**) Temperature dependent of resistivity (top, ref.*16*) and scattering intensity of (0.3,0,1) stripe (bottom, ref.*13*) of x=0.4 compound. (**E**) A schematic Fermi surface plot of $La_{2-2x}Sr_{1+2x}Mn_2O_7$ shows hole-like bonding (solid green) and antibonding (dashed green) portions of bilayer-split bands, as well as the electron-like zone center pocket. The arrows indicate two types of nesting vectors. As discussed in the text, they correspond to "**V**ertical" (red) and "**D**iagonal" (blue) stripes.

Fig. 2.

Doping dependence of Fermi surface nesting vectors of $La_{2-2x}Sr_{1+2x}Mn_2O_7$. (**A**) Antinodal MDCs at the Fermi level for various doping levels showing a non-monotonic evolution with doping. (**B**) Comparison of experimental data and theoretical calculations. Band theory results are shown in triangles with a hatched area indicating the general trend. Down-triangles are from ref.*22*; up-triangles are our own calculations as discussed in the supporting material. Filled green circles are ARPES data of the bonding bands or the non-bilayer split bands. "V"s and "D"s represent the "vertical stripe" and "diagonal stripe" ordering from scattering experiments (*6,7,8,9,10,11,12,13,14,15*). The grey shaded area shows the experimental trend determined by ARPES and x-ray scattering.

Fig 3.

Electronic structure of $La_{2-2x}Sr_{1+2x}Mn_2O_7$ (x=0.4) below and above $T_c$=120K. (**A,B**) Stacked EDCs along the blue cut in the inset, taken at 150 K and 50 K, respectively, showing the change of spectral weight. In the inset of panel (A), only the electron-like pocket and hole-like bonding pocket are shown. (**C-F**) Spectral intensities at constant energy across the first Brillouin zone. Panels (C,D) are at 150K and (E,F) at 50K, while (C,E) are at $E_F$ and (D,F) are at 0.2 eV below $E_F$. Panels (C, D, and E) have an identical gray scale while the spectral weight in panel (F) was scaled down by a factor of 4 to prevent saturation.

**Supporting material:**

**Experimental Details:**

The single crystals were grown using the traveling-solvent floating zone method as described elsewhere (*S1*). Our experiments were performed at beamlines 7.0.1, 10.0.1, and 12.0.1 of the Advanced Light Source, Berkeley using Scienta electron spectrometers. All data shown here were taken in a vacuum better than $3\times10^{-11}$ torr. Bilayer manganite samples with doping levels from x=0.36 to 0.6 (see Fig. 1 for the doping phase diagram) were cleaved and measured at 20 K. We took advantage of our recent discovery of, and ability to deconvolve, bilayer splitting in these materials (*S2*), enabling much more careful studies of the Fermi surface nesting vectors.

**Band theory computation:**

Detailed calculations of the electronic structure and magnetic configurations of the x=0.5 hole-doped double layered manganite $LaSr_2Mn_2O_7$ were performed within the all-electron full-potential KKR and LAPW methods (*S3,S4*). We calculated the electronic structure of both ferromagnetic and A-type antiferromagnetic LaSr$_2$Mn$_2$O$_7$. The results for the ferromagnetic case are in accord with our previous studies (*S5*). The electronic structure for the x=0.4 or x=0.6 compounds was then obtained by using a rigid band model in which the Fermi energy was adjusted to accommodate the correct number of electrons in the ferromagnetic or A-type antiferromagnetic case. Very similar results are obtained if the virtual crystal approximation (VCA) is used to model the effects of La/Sr disorder (*S6*). To model exchange-correlation effects we have considered the Local Spin Density Approximation (LSDA), the Generalized Gradient Approximation (GGA), and LSDA+U methods. For the LSDA+U calculations, we have used the Coulomb parameter value of U=7.2 eV suggested by Medvedeva et al. (*S7*). The crystal parameters were taken from Ref. *S8*.

**The discrepancy between theory and experiment**

The fact that the two completely different types of spectroscopies (scattering and ARPES) give such a close agreement indicates to us that they must be returning the correct values, and that the band calculation misses some ingredients. Such a failure of the band calculations to get the correct evolution of the Fermi surface with doping is highly unusual. We have considered whether the exchange splitting might be suppressed due to screening by more conduction electrons at lower hole doping such that electrons fill in the minority bands faster than the rigid band model predicts as hole doping decreases. We attempted to test this within the local spin density approximation (LSDA) and the generalized gradient approximation (GGA), using the virtual crystal approximation (VCA) to go beyond the rigid band filling. Although some changes in the band structures are found, the overall doping evolution is similar to the case of rigid band filling and fails to explain the non-monotonic behavior seen in experiments. These results are in general consistent with Ref. *S9* which failed to show the non-monotonic doping dependence shown here.

A couple of options are available to explain the disagreement found here. With the presence of dynamic stripes, one has to do with the strong (and potentially intrinsic) inhomogeneity in these and other correlated electron systems, whereas the band calculations were done for a homogeneous system. Theoretical calculations have shown that such inhomogeneity may have strong and unexpected consequences on the electronic structure (*S10*). Additionally, orbital degrees of freedom may play an important role in shifting spectral weight from one set of orbitals to another. Band calculations suggest that the proportion between $d_{x^2-y^2}$ and $d_{3z^2-r^2}$ states in these relevant bands varies with energy and momentum (*S9,S11*). The antibonding band is primarily of $d_{x^2-y^2}$ character, while the mixing of $d_{x^2-y^2}$ and $d_{3z^2-r^2}$ states in the bonding band is significant (*S9,S11*). The electron-like piece of Fermi surface is mainly of $d_{3z^2-r^2}$ character, but with a small portion of $d_{x^2-y^2}$ states near $E_F$ (*S9,S11*). Our preliminary ARPES measurements (unpublished) show that the Fermi surface centered at the Γ point, consisting of primarily $d_{3z^2-r^2}$ orbital states, vanishes more quickly with increasing doping than is predicted by theoretical calculations, with this in fact helping to counter the trends observed for the bonding Fermi surface. Other experiments (*S12,S13*) also suggest that, for x<0.5, doping hole carriers removes electrons mainly of $d_{3z^2-r^2}$ character, while the electron count of $d_{x^2-y^2}$ orbital is less modified. Such a possibility suggests that the additional $d_{3z^2-r^2}$ orbital can act as a charge "reservoir" or "lever" to enable the bonding band to follow tendencies or preferences other than those expected from the rigid doping model in band calculations.

S1. J. F. Mitchell *et al.*, *Phys. Rev. B* **55**, 63 (1997).

S2. Z. Sun *et al.*, *Phys. Rev. Lett.* **97**, 056401 (2006).

S3. A. Bansil, S. Kaprzyk, P. E. Mijnarends, J. Tobola, *Phys. Rev. B* **60**, 13396 (1999).

S4. P. Blaha, K. Schwarz, G. K. H. Madsen, D. Kvasnicka and J. Luitz, WIEN2k, An Augmented Plane Wave + Local Orbitals Program for Calculating Crystal Properties (Karlheinz Schwarz, Techn. Universität Wien, Austria), 2001. ISBN 3-9501031-1-2.

S5. P. E. Mijnarends *et al.*, *Phys. Rev. B* **75**, 014428 (2007).

S6. H. Lin, S. Sahrakorpi, R. S. Markiewicz, A. Bansil, *Phys. Rev. Lett.* **96**, 097001 (2006).

S7. J. E. Medvedeva *et al.*, *Journal of Magnetism and Magnetic Materials* **237**, 47 (2001).

S8. R. Seshadri *et al.*, *Solid State Commun.* **101**, 453 (1997).

S9. R. Saniz, M. R. Norman, A. J. Freeman, *Phys. Rev. Lett.* **101**, 236402 (2008).

S10. E. Dagotto, *Science* **309**, 257 (2005).

S11. Z. Sun *et al.*, *Phys. Rev. B* **78**, 075101 (2008).

S12. T. Kimura, Y. Tokura, *Annu. Rev. Mater. Sci.* **30**, 451 (2000).

S13. A. Koizumi *et al.*, *Phys. Rev. Lett.* **86**, 5589 (2001).

Fig. 1

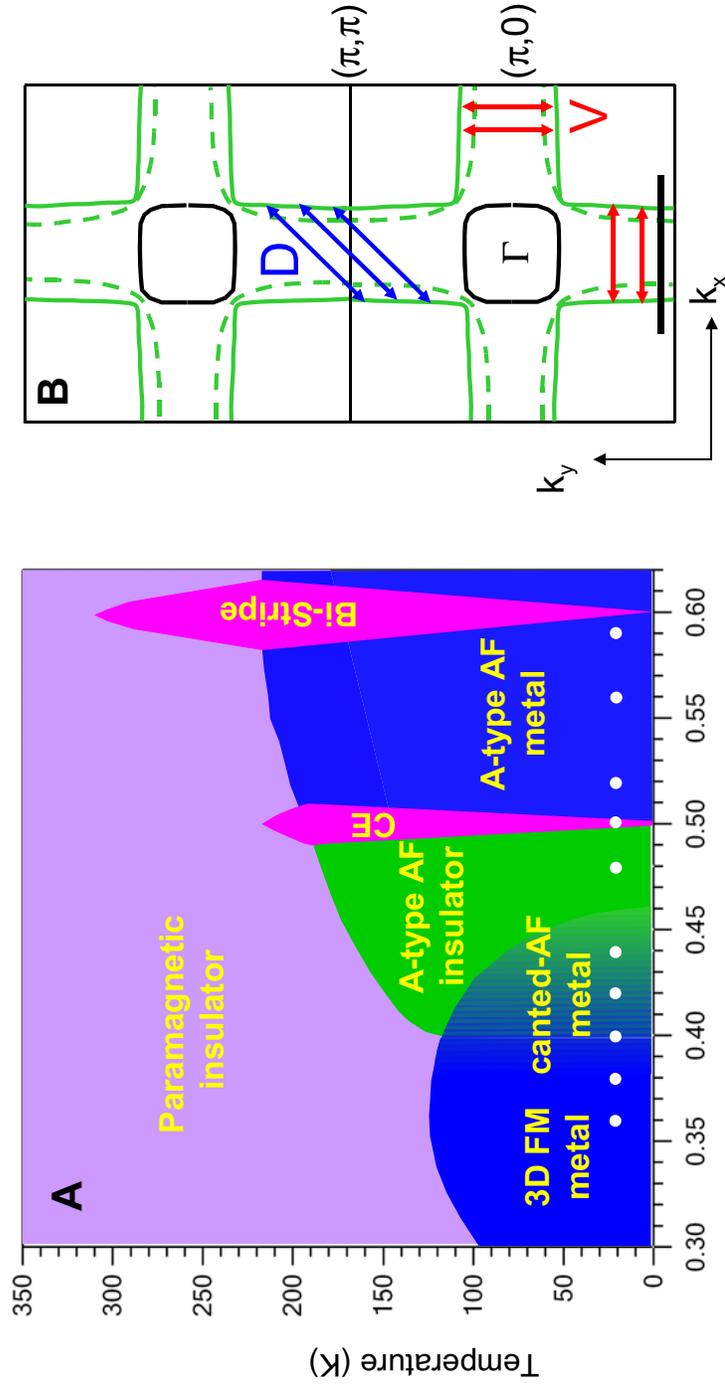
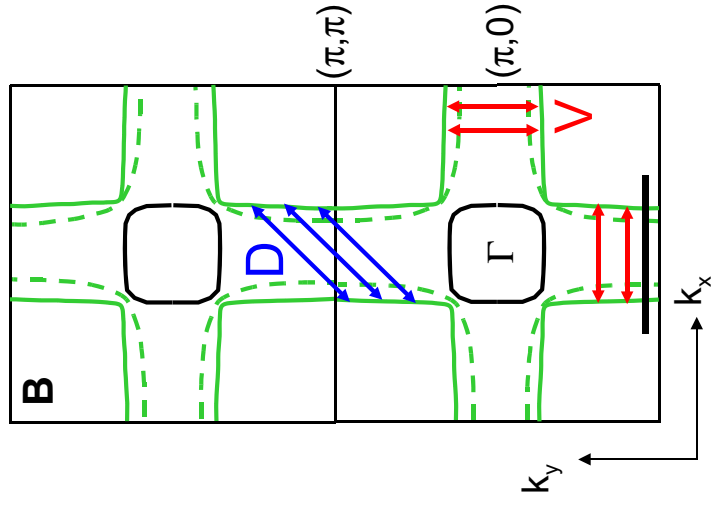
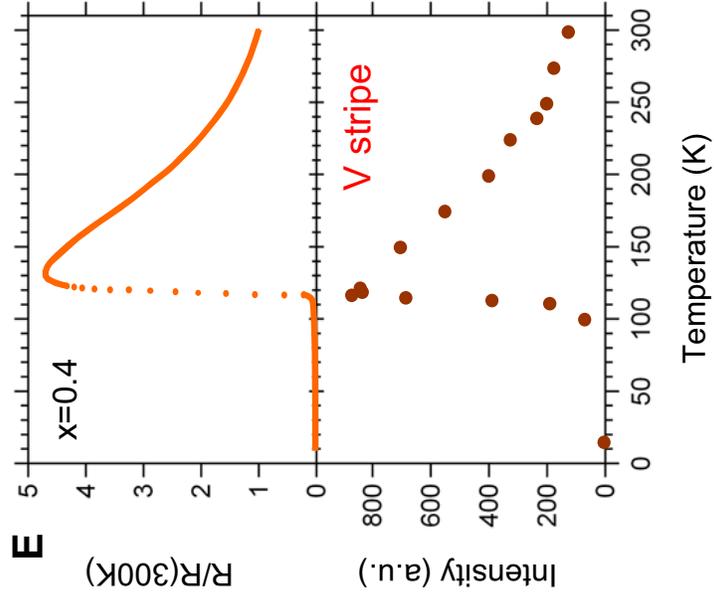
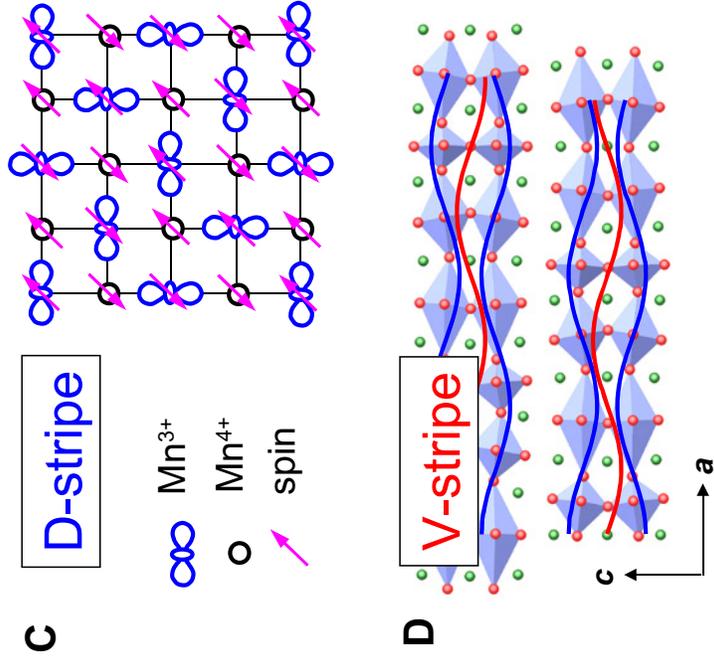

Fig. 2

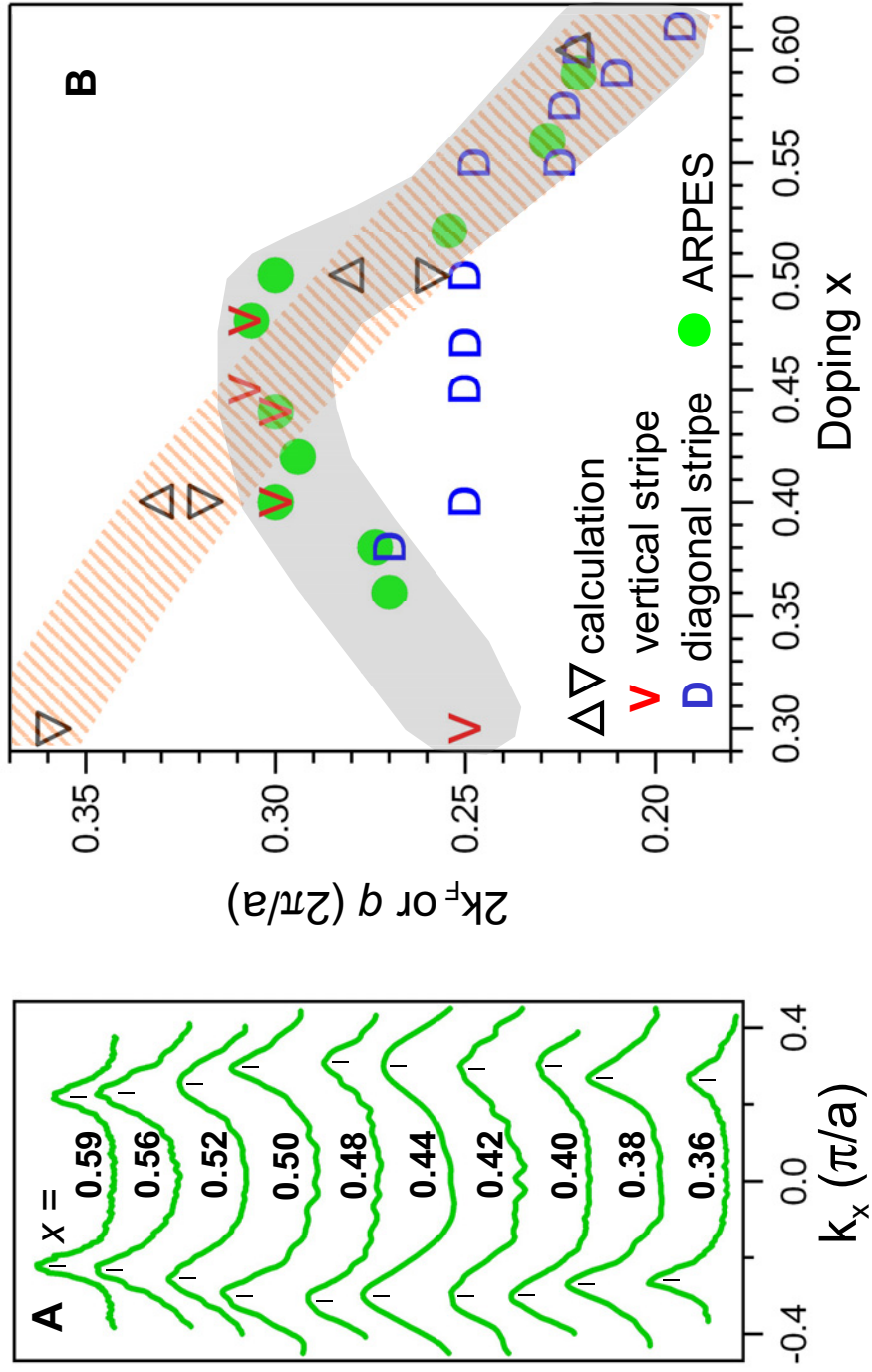

Fig. 3

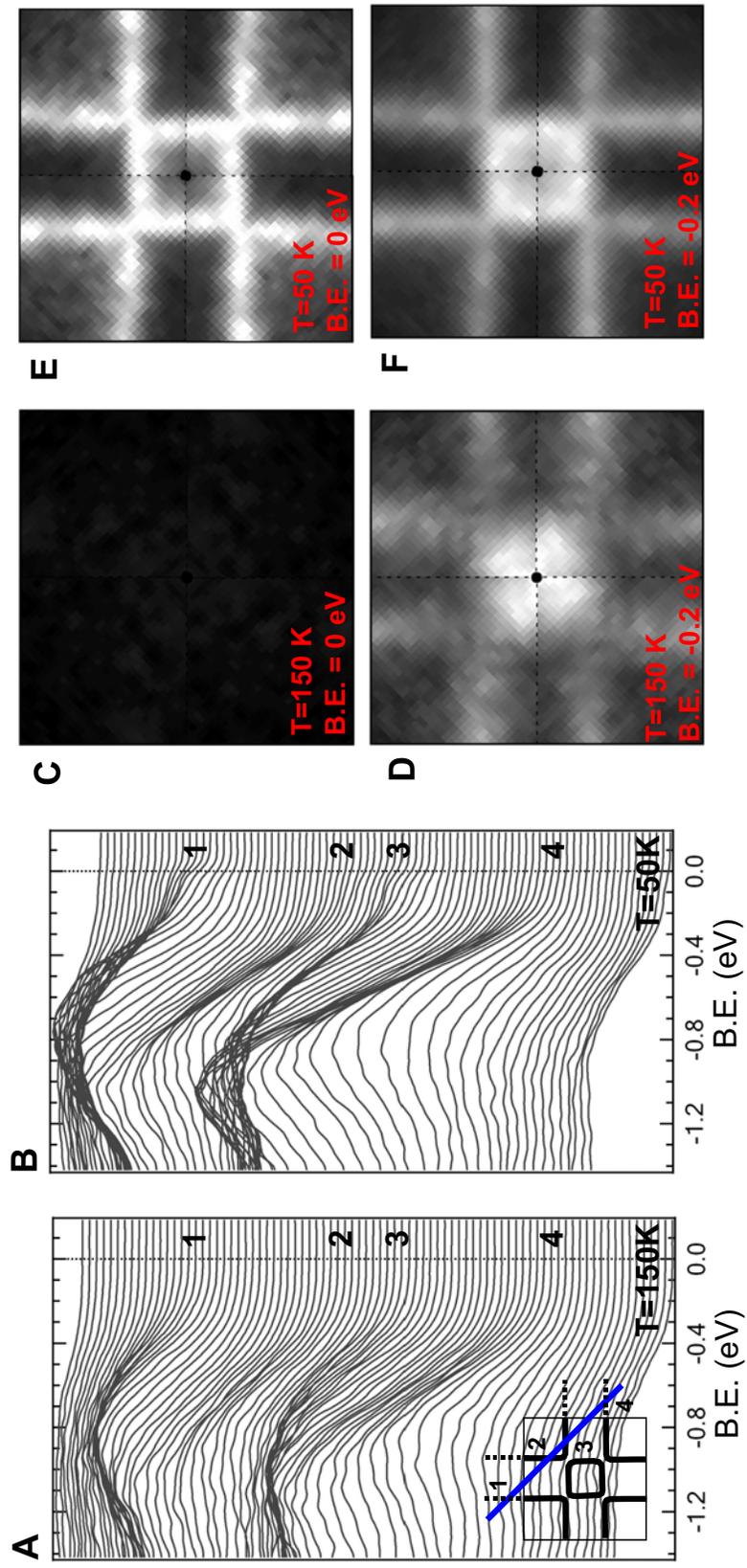